\documentclass[aoas,nameyear,seceqn,MSNbibl,dvips]{arximspdf}
\usepackage{graphicx}

\doi{10.1214/10-AOAS448}
\volume{5}
\issue{2B}
\pubyear{2011}
\firstpage{1586}
\lastpage{1610}

\makeatletter
\newtheorem{prop}{Proposition}
\let\epsilon\varepsilon
\makeatother

\begin{document}
\begin{frontmatter}

\title{Degradation modeling applied to residual lifetime prediction
using functional data analysis}
\runtitle{Degradation modeling using functional data analysis}

\begin{aug}
\author[A]{\fnms{Rensheng R.} \snm{Zhou}\thanksref{m1}\ead[label=e1]{rzhou8@gatech.edu}},
\author[A]{\fnms{Nicoleta} \snm{Serban}\corref{}\ead[label=e2]{nserban@isye.gatech.edu}}
and
\author[A]{\fnms{Nagi} \snm{Gebraeel}\thanksref{m1}\ead[label=e3]{nagi@isye.gatech.edu}}
\thankstext{m1}{Supported in part by the NSF Grant CMMI-0738647.}
\runauthor{R. R. Zhou, N. Serban and N. Gebraeel}
\affiliation{Georgia Institute of Technology}
\address[A]{Industrial and Systems Engineering\\
Georgia Institute of Technology\\
765 Ferst Drive, NW\\
Atlanta, Georgia 30332-0205\\
USA\\
\printead{e1}\\
\hphantom{E-mail: }\printead*{e2}\\
\hphantom{E-mail: }\printead*{e3}} 
\end{aug}

\received{\smonth{5} \syear{2010}}
\revised{\smonth{10} \syear{2010}}

%
\begin{abstract}
Sensor-based degradation signals measure the
accumulation of damage of an engineering system using sensor
technology. Degradation signals can be used to estimate, for
example, the distribution of the remaining life of partially
degraded systems and/or their components. In this paper we present
a nonparametric degradation modeling framework for making inference
on the evolution of degradation signals that are observed sparsely
or over short intervals of times. Furthermore, an empirical Bayes
approach is used to update the stochastic parameters of the
degradation model in real-time using training degradation signals
for online monitoring of components operating in the field. The
primary application of this Bayesian framework is updating the
residual lifetime up to a degradation threshold of partially
degraded components. We validate our degradation modeling approach
using a real-world crack growth data set as well as a case study of
simulated degradation signals.
\end{abstract}

%

\end{frontmatter}

\section{Introduction}

Most failures of engineering systems result from a gradual and
irreversible accumulation of damage that occurs during a system's
life cycle. This process is known as \textit{degradation} [\citet
{BogKoz85}]. In many applications, it can be very difficult
to assess and observe physical degradation, especially when
real-time observations are required. However, degradation
processes are almost always associated with some manifestations
that are much easier to observe and monitor overtime. Generally,
the evolution of these manifestations can be monitored using
sensor technology through a process known as \textit{Condition
Monitoring} (CM). The observed condition-based signals are known
as \textit{degradation signals} [\citet{Nel90}] and are usually
correlated with the underlying physical degradation process. Some
examples of degradation signals include vibration signals for
monitoring excessive wear in rotating machinery, acoustic
emissions for monitoring crack propagation, temperature changes
and oil debris for monitoring engine lubrication and many others.

\begin{figure}[b]

\includegraphics{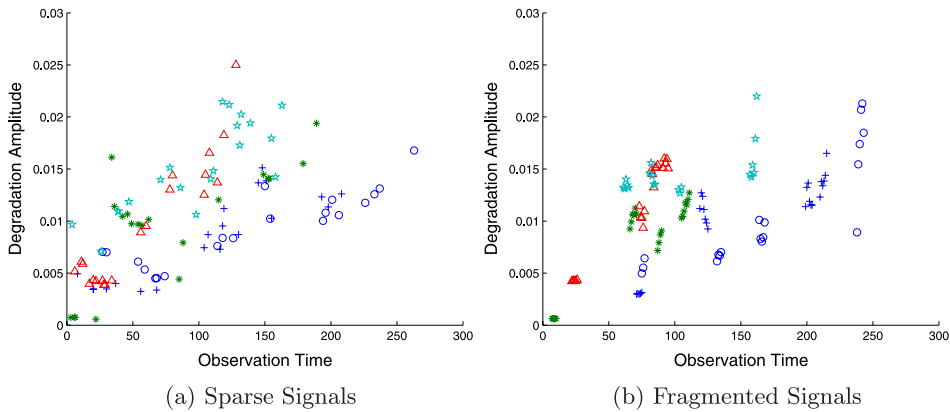}

\caption{Examples of incomplete degradation signals.} \label{fig:ex:signals}
\end{figure}

\textit{Degradation modeling} attempts to characterize the evolution of
degradation signals. There is a significant number of research
works that have focused on degradation models; these include models
presented in \citet{LuMee93}, \citet{PadTom04},
Gebraeel et al. (\citeyear{Gebetal05}), M\"{u}ller and Zhang (\citeyear{MulZha05}), \citet{Geb06} and
\citet{ParPad06}. Many of these models rely on a
representative sample of {complete degradation signals}. A \textit
{complete degradation signal} is a continuously observed signal that
captures the degradation process of a component from a brand ``New
State'' to a completely ``Failed State.''

Unfortunately, building a database of complete degradation signals
can be very expensive and time consuming in applications, such as
monitoring of jet engines, turbines, power generating units,
structures and bridges and many others. For example, in
applications consisting of relatively static structures such as
bridges, degradation usually takes place very slowly (several tens
of years). Since the system is relatively static, it suffices to
observe the degradation process at intermittent discrete time
points. The result is a~sparsely observed degradation signal such
as the signals depicted in Figure~\ref{fig:ex:signals}(a). In
contrast, in applications consisting of dynamic systems, such as
turbines, generators and machines, degradation cannot be reasonably
assessed by sparse measurements. At the same time, continuous
observations of the complete degradation process of such systems are
economically unjustifiable. Usually, the only way to gain a
relatively accurate understanding of the health/performance of a
dynamic system is to monitor its performance over a time interval.
In naval maritime applications, power generating units of an
aircraft are removed, tested for a short period of time (during
which degradation data can be acquired) and put back into operation.
The result is a collection of fragmented degradation signals as
depicted in Figure \ref{fig:ex:signals}(b).

In this paper we develop a degradation model that applies to
incomplete degradation signals as well as complete degradation
signals. An \textit{incomplete degradation signal} is defined as a
signal that consists of sparse observations of the degradation
process or continuous observations made over short time intervals
(fragments). \textit{One challenge in such applications is that the
evolution of the degradation signals cannot be readily assessed to
determine the parametric form of the underlying degradation model.}
This is because one cannot clearly trace how a degradation signal
progresses over time from incomplete observations. For example, is
there a well defined parametric model that describes the signals in
Figure \ref{fig:ex:signals}?

To overcome this challenge, the underlying degradation model in this
paper is assumed nonparametric. Most degradation models used to
characterize the evolution of sensor-based degradation signals are
parametric models. A~common approach is to model the degradation
signals using a parametric (linear) model with random coefficients
[\citet{LuMee93}; Gebraeel et al. (\citeyear{Gebetal05}); \citet{Geb06}]. Other
modeling approaches assume that the degradation signal follows a
Brownian motion process [\citet{DokHoy92}; \citet{PetYou99}]
or a Gaussian process with known covariance structure [\citet
{PadTom04}; \citet{ParPad06}]. In contrast, we assume
that the mean and covariance functions of the degradation process
are unknown and they are estimated based on an assembly of training
incomplete degradation signals. The mean function is estimated using
standard nonparametric regression methods such as local smoothing
[\citet{FanYao03}]. The covariance function is decomposed using the
Karhunen--Lo\`{e}ve decomposition [\citet{Kar}; Lo\`{e}ve (\citeyear{Loe45})]
and estimated using the Functional Principal Component Analysis
(FPCA) method introduced by \citet{YaoMulWan05}.

Under the nonparametric modeling framework, one condition for
accurate estimation of the mean and covariance functions is that the
degradation process is densely observed throughout its support.
However, in applications where the degradation signals are
incompletely sampled, not all degradation signals are observed up to
the point of failure; in addition, only a few components will
survive up to the maximum time point of the degradation process
support. Consequently, the degradation process is commonly
under-sampled close to the upper bound of its support. To overcome
this difficulty, we introduce a nonuniform sampling procedure for
collecting incomplete degradation signals, which ensures relatively
dense coverage throughout the sampling time domain.

One important application of degradation modeling is predicting the
lifetime of components operating in the field. For real-time
monitoring, an empirical Bayes approach is introduced to update the
stochastic parameters of the degradation model. In this paper we
focus on estimation of the distribution of the residual life up to a
degradation threshold for partially degraded components using
training degradation signals which are sparsely or completely
observed. Other applications of the degradation modeling and the
Bayesian updating procedure are estimation of the lifetime at a
specified degradation level and estimation of the degradation level
at a specified lifetime.

We evaluate the performance of our methodology using both a crack
growth data set and simulated degradation signals. In these empirical
studies we compare parametric to nonparametric degradation modeling,
assess the estimation accuracy of the remaining lifetime for complete
and incomplete signals, and contrast uniform vs. nonuniform sampling
procedures for acquiring ensembles of incomplete degradation signals.
In both studies there is not a significant decrease in the accuracy of
the residual life estimation when using ensembles of incomplete instead
of complete signals. We also highlight the robustness of our approach
by comparing it with misspecified parametric models, which are common
when the underlying degradation process is complicated and sparsely
observed. Last, we show in the simulation study that using a nonuniform
sampling procedure that ensures dense observation of the sampling time
domain reduces the estimation error. Based on these empirical studies,
we conclude that the nonparametric approach introduced in this paper is
efficient in characterizing the underlying degradation process and it
is more robust to model misspecification than parametric approaches,
which is particularly important when the training signals are
incompletely observed (sparse or fragmented).

The remainder of the paper is organized as follows. Section
\ref{sec:degradation:model} discusses the development of our
degradation modeling framework. The empirical Bayes approach for
updating the degradation distribution of a partially degraded
component is introduced in Section \ref{sec:update:distr}. The
derivation of the remaining lifetime distribution under the
empirical Bayes approach is presented in Section~%
\ref{sec:lifetime:distr}. In Section \ref{sec:expt:design} we
introduce an experimental design for sampling incomplete degradation
signals. We discuss performance results of our methodology using
real-world and simulated degradation signals in Section
\ref{sec:crack:study} and \ref{sec:simulation:study}, respectively.

\section{Sensor-based degradation modeling}\label{sec:degradation:model}

We denote the observed degradation signals $S_i(t_{ij})$, for
$j=1,\ldots,m_i$ ($m_i$ is the number of observation time points
for signal $i$) and $i=1,\ldots,n$ ($n$ is the number of signals)
where $\{t_{ij}\}_{j=1,\ldots,m_i}$ are the observation time
points in a bounded time domain $[0,M]$ for signal $i$. Note that
$M$ will always be finite since any industrial application has a
finite time of failure. We model the distribution of the signals
$S_i(t)$ by borrowing information across multiple degradation
signals. We decompose the degradation signal as
%
\begin{equation}\label{eq:model}
S_i(t) = \mu(t)+ X_i(t)+\sigma\epsilon_i(t),
\end{equation}
where $\mu(t)$ is the underlying trend of the degradation process
and is assumed to be fixed but unknown, and $X_i(t)$ represents the random
deviation from the underlying degradation trend. We also assume
$X_i(t)$ and $\epsilon_i(t)$ are independent.

The model in (\ref{eq:model}) is a general decomposition for
functional data with various modeling alternatives and assumptions
for the model components: $\mu(t)$, $X_i(t)$, and $\epsilon_i(t)$.
In this paper we discuss one such modeling alternative which
applies to sparse and fragmented signals as well as to complete
signals and it applies under the assumption that the observation
time points $\{t_{ij}\}_{j=1,\ldots,m_i}$ are fixed but not
necessarily equally spaced and the assumption that the error terms
$\epsilon_i(t)$ are independent and identically distributed.
Deviations from these assumptions may require some modifications to
the modeling approach discussed in this paper.

In our modeling approach, the degradation signal $S_i(t)$ follows a
stochastic process with mean $\mu(t)$ and stochastic deviations
$X_i(t)$ with mean zero and covariance $\operatorname{cov}(t,t')$. The mean
function $\mu(t)$ and the covariance surface $\operatorname{cov}(t,t')$ are
both assumed to be nonparametric, that is, no prespecified assumption
on their shape. This generalized assumption encompasses the
particular cases developed earlier by Gebraeel et al. (\citeyear{Gebetal05})
and \citet{Geb06}, which assume a linear trend, $\mu(t) =
\alpha+\beta t$ where $\alpha\sim N(0,\delta_\alpha)$ and $\beta\sim
N(0,\delta_\beta)$, and parametric covariance structure
$\operatorname{cov}(t,t') = \delta_\alpha+\delta_\beta tt'$.

The following steps discuss how we estimate the mean function and
the covariance surface of our degradation model.
\begin{longlist}[\textit{Step} 1:]
\item[\textit{Step} 1:] We use nonparametric methods to estimate the mean $\mu
(t)$. In this paper we use local quadratic smoothing [\citet
{FanYao03}] to allow estimation of the mean function under general
settings including complete and incomplete (sparse and fragmented)
signals. The bandwidth parameter, which controls the smoothing level,
is selected using the leave-one-curve-out cross-validation method
[\citet{RicSil91}]. Alternative estimation methods include
decomposition of the mean function using a basis of functions (e.g.,
splines, Fourier, wavelets) and estimate the coefficients using
parametric methods. These alternative methods will apply under various
signal behaviors (e.g., smooth vs. with sharp changes, uniformly vs.
nonuniformly sampled).

\item[\textit{Step} 2:] The covariance surface is estimated using the demeaned
data, $S_i(t)-\hat\mu(t)$, where $\hat\mu(t)$ is the local quadratic
smoothing estimate of $\mu(t)$. Using the Karhunen--Lo\'{e}ve
decomposition [\citet{Kar}; Lo\`{e}ve (\citeyear{Loe45})], the covariance,
$\operatorname{cov}(t,t') = \operatorname{Cov}(S_i(t),S_i(t'))$, can be expressed as
follows:
%
\begin{equation}\label{eq:cov}
\operatorname{cov}(t,t') = \sum_{k=1}^{\infty} \lambda_k\phi_k(t)\phi_k(t'),\qquad
t,t'\in[0,M],
\end{equation}
where $\phi_k(t)$ for $k=1,2,\ldots$ are the associated
eigenfunctions with support $[0,M]$ and $\lambda_1\geq
\lambda_2\geq\cdots$ are the ordered eigenvalues. Based on this
decomposition, the deviations from the underlying degradation
trend $X_i(t)$ are decomposed using the following expression:
%
\begin{equation}\label{eq:decomp}
X_i(t_{ij}) =\sum_{k=1}^{\infty} \xi_{ik} \phi_k(t_{ij}),
\end{equation}
where $\xi_{ik}$ called \textit{scores} are uncorrelated random
effects with mean zero and variance $\mathbb{E}(\xi_{ik}^2) =
\lambda_k$. The decomposition in equation (\ref{eq:decomp}) is an
infinite sum. Generally, only a small number of eigenvalues are
commonly significantly nonzero. For the eigenvalues which are
approximately zero, the corresponding scores will also be
approximately zero. Consequently, we use a truncated version of
this decomposition. Therefore, expression (\ref{eq:decomp}) can be
approximated as follows:
%
\begin{equation}\label{eq:decomp:red}
X_i(t_{ij}) =\sum_{k=1}^{K} \xi_{ik} \phi_k(t_{ij}),
\end{equation}
where $K$ is the number of significantly nonzero eigenvalues. We
select $K$ to minimize the modified Akaike criterion defined by
\citet{YaoMulWan05}.
\end{longlist}

In the statistical literature this method has been coined
Functional Principal Component Analysis (FPCA). The key reference
for FPCA is Ramsay and Silverman [(\citeyear{RamSil05}), Chapter 8]. Another important
reference is \citet{YaoMulWan05}, in which the authors derived
theoretical results
for model parameter consistency and asymptotic ($n$ large)
distribution results under the assumption that the scores follow a
normal distribution.

An alternative method for estimating the covariance function of the
process $X_i(t)$ is decomposing the covariance function as in equation
(\ref{eq:cov}) where the basis of functions $\{\phi_k,  k= 1,2, \ldots
\}$ is fixed [\citet{JamHasSug00}]. However, this approach doesn't
allow dimensionality reduction in the same way FPCA does and it is not
theoretically founded.

\section{Degradation model updating}\label{sec:update:distr}

Next, we consider a component operating in the field called \textit
{fielded component}. Assume that we have observed its degradation
signal at a vector of time $\mathbf{t}=(t_1,\ldots,t_{m^*})$; therefore,
$S(\mathbf{t})$~denotes the observed signal of the testing component, $m^*$
represents the number of observations and $t^*=t_{m^*}$ denotes the
latest observation time. In this section we introduce an Empirical
Bayes approach which allows real-time updating of the distribution of
the degradation process for partially degraded components given the
observed signal $S(\mathbf{t})$ and the prior distribution of the scores $\xi
_{ik}$ for $k=1,2,\ldots.$ The prior distribution of the scores is
estimated empirically from a set of historical degradation signals.

Proposition \ref{prop1} illustrates the updating procedure assuming that the
prior distribution of the scores is normal and assuming that the mean
function~$\mu(t)$ and the basis of functions $\phi_k(t)$, $k=1,\ldots,K$,
are fixed. The proof of this proposition
follows from the theory of Bayesian linear models.

\begin{prop}\label{prop1}
Assume that $S(t)$ follows
\[
S(t) = \mu(t)+\sum_{k=1}^{K} \xi_{k} \phi_k(t)+\epsilon(t),
\]
where the prior distribution of $\xi_{k}$ is $N(0,\lambda_k)$ with $\xi
_{1}, \ldots, \xi_{K}$ uncorrelated; $\epsilon(t)$ are independent of
$\xi_{k}$ for $k=1,\ldots,K$; the distribution of $\epsilon(t)$ is
$N(0,\sigma^2)$ with $\sigma^2$ fixed. It follows that the posterior
distribution of the scores is
\[
(\xi_{1}^*, \ldots, \xi_{K}^* )'\sim N(Cd,C),
\]
where
\[
C = \biggl(\frac{1}{\sigma^2}P(\mathbf{t})'P(\mathbf{t}) + \Lambda^{-1}
\biggr)^{-1} \quad \mbox{and} \quad  d=\frac{1}{\sigma^2}P(\mathbf{t})'\bigl(S(\mathbf{t})-\mu
(\mathbf{t})\bigr)
\]
with
\begin{eqnarray}\label{eq:xdef}
S({\mathbf{t}})&=&(S(t_1),\ldots,S(t_{m^*}))',\qquad
\mu(\mathbf{t})=(\mu(t_1),\ldots,\mu(t_{m^*}))',\nonumber\\[-8pt]\\[-8pt]
\Lambda&=&\operatorname{diag}(\lambda_1,\ldots,\lambda_K),\qquad  P(\mathbf{t})=
\pmatrix{
\phi_1(t_1) & \ldots& \phi_K(t_1) \cr
\ldots& \ldots& \ldots\cr
\phi_1(t_{m^*}) & \ldots& \phi_K(t_{m^*})
}.\nonumber
\end{eqnarray}
\end{prop}

In Proposition \ref{prop1} the prior distribution of the scores is specified by
the variance parameters $\lambda_k$, $ k=1,\ldots,K$, which are
estimated using the degradation model in Section \ref
{sec:degradation:model} and based on a set of incomplete or complete
training degradation signals. Specifically, we first apply Functional
Principal Component Analysis on the historical degradation signals
which will further provide estimates for the variance parameters
$\lambda_k$, $ k=1,\ldots,K$, and the eigenfunctions $\phi_k$, $
k=1,\ldots,K$. Based on these estimates, we obtain the posterior
distributions of the updated scores $\xi_{1}^*, \ldots, \xi_{K}^*$
since the matrix $C$ and the vector $d$ are fully determined by the
eigenvalues $\lambda_k$, $k=1,\ldots,K$, and the eigenfunctions $\phi
_k$, $ k=1,\ldots,K$. The expectation of the posterior scores is nonzero
and, therefore, we denote the posterior mean function $\mu^*(t) = \mu
(t)+\sum_{k=1}^{K} E(\xi_{k}^*) \phi_k(t)$.

Following Proposition \ref{prop1}, the expectation of the posterior distribution
follows the same formula as the conditional expectation estimator in
\citet{YaoMulWan05}, equation (4). Generally, this similarity applies
under the empirical Bayesian prior derived from FPCA. On the other
hand, the sampling distribution of the conditional expectation
estimator in \citet{YaoMulWan05} is different from the posterior
distribution of $\xi^*_k, k=1,\ldots,K$, since their variances are not
equal. Moreover, the conditional expectation estimator and its mean
estimation error in \citet{YaoMulWan05} are conditional on the training
observations, whereas the posterior distribution in Proposition \ref{prop1} is
conditional on the observations of a new component.

The advantage of this Bayesian framework is that it unifies the
conditional expectation estimation and prediction into a procedure
which allows updating the distribution of the degradation process
for a new component. We can therefore use the posterior distribution
of the scores for a partially degraded signal to estimate the
distribution of various statistical summaries, including the lifetime
at a specified degradation level and estimation of the degradation
level at a specified time. In the next section we discuss one
specific application to this updating framework: residual life
estimation.

\section{Remaining life distribution}\label{sec:lifetime:distr}

In this paper we focus our attention on engineering
applications where a soft-failure of a system occurs once its underlying
degradation process reaches a predetermined critical threshold. This
critical threshold is commonly used to initiate maintenance
activities such as repair and/or component replacement well in
advance of catastrophic failure. Consequently, degradation data can
still be observed beyond the critical threshold. In this section we
describe how our degradation modeling framework is applied to
estimate the distribution of remaining life up to a~degradation
threshold of partially degraded systems.

In the remainder of this section, $S^*(\cdot)$ will denote the
underlying degradation process of a partially degraded system. Based on
the degradation process $S^*(\cdot)$, the failure time of a system is
defined as
%
\begin{equation}\label{eq:lifetime}
T = \inf_{t\in[0,M]}\{S^*(t)\geq D\}.
\end{equation}

One has to bear in mind that $T$ may not exist if the threshold $D$ is
set too high, that is, the component may fail before its degradation
signal reaches the threshold. The selection of the failure threshold
$D$ is an important problem, however, this aspect is beyond the scope
of this paper. In this work, we assume that $T$ exists, and the
threshold $D$ is known a priori. This is a reasonable assumption
because in many industrial applications failure/alarm thresholds are
usually based on subjective engineering judgement or well-accepted
standards, such as the International Standards Organization (ISO)
(e.g., the ISO 2372 is used for defining acceptable vibration
threshold levels for different machine classifications). A second
assumption is that the failure time $T$ is smaller than a
maximum failure time $M$. This assumption is also reasonable, as in
practice a component may be replaced after a given period of time even
if it did not fail.

The distribution of the residual life (RLD) of a partially degraded
component at a fixed time $t^*\in[0,M]$ is estimated assuming that the
degradation process $S^*(\cdot)$ of the component follows a posterior
distribution based on Proposition \ref{prop1}. We estimate the distribution of
the residual life (RLD) using
\[
R(y|t^*) = P\bigl(T-t^*\leq y  | S^*(t)\sim\operatorname{Gaussian}(\mu^*(t), \operatorname{Cov}^*(t,t')),  t^*\leq T\leq M\bigr),
\]
where $\mu^*(t)$ and $\operatorname{Cov}^*(t,t')$ are the posterior mean and
covariance functions of the degradation process $S^*(\cdot)$. The
derivations of $\mu^*(t)$ and $\operatorname{Cov}^*(t,t')$ are based on the
results of Proposition \ref{prop1}. We note here that the distribution of the RLD
above is not conditional on the observed signal of the partially
degraded component but on the posterior distribution of its degradation
process; since the degradation is only partially observed and most
often sparsely sampled, conditioning on the posterior distribution will
generally provide a more accurate RLD estimator since we incorporate
the additional information in the training degradation signals.

\begin{figure}

\includegraphics{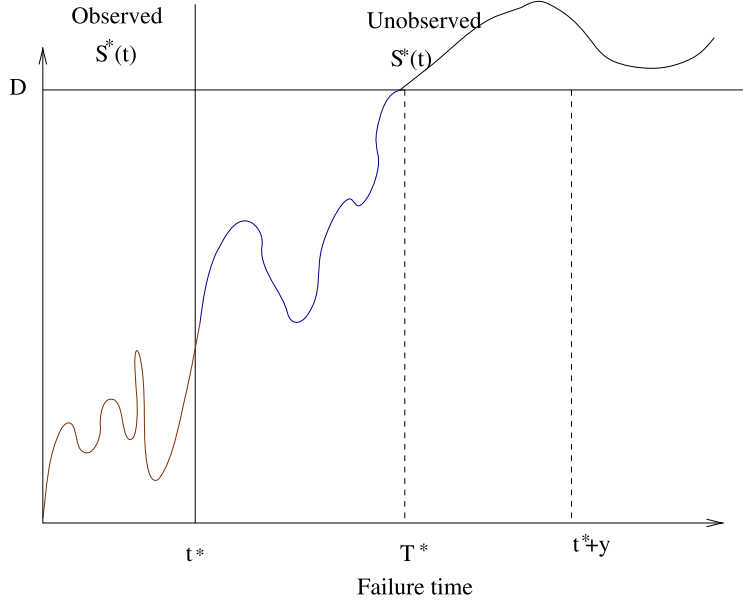}

\caption{Example of a signal for which assumption \textup{A.2} holds.}
\label{fig:exp:assumption:A2}
\end{figure}

Furthermore, we estimate RLD under two assumptions:
\begin{enumerate}[A.0]
\item[A.1] The new component has not failed up to the last observation
time point $t^*$, that is, the failure time becomes
\[
T = \inf_{t\in[0,M]} \{S^*(t)\geq D\} = \inf_{t\in[t^*,M]} \{
S^*(t)\geq D\}:=T^*.
\]

\item[A.2] We assume the probability that the degradation process
$S^*(t)$ crosses
back the threshold $D$ after the failure time $T^*$ is negligible, that
is, $P(S^*(T^*+y)<D) \approx0$ for all $y>0$. This implies, if we
condition on $y\ge T^*-t^*>0$, which is the same as conditioning on
$T^*\le t^*+y$, $P(S^*(t^*+y)<D|T^*\leq t^*+y) \approx0$. This further implies
\begin{eqnarray*}
P\bigl(S^*(t^*+y)\geq D\bigr) &=& P\bigl(S^*(t^*+y) \geq D|T^*\leq t^*+y\bigr) P(T^*\leq
t^*+y)\\
&\approx& P(T^*\leq t^*+y).
\end{eqnarray*}
\end{enumerate}
Under these two assumptions, the RLD becomes
\[
R(y|t^*) = \mbox{(by A.1)} \  P\bigl(T^*-t^*\leq y|  S^*(t)\bigr) \approx\mbox
{(by A.2)} \ P\bigl(S^*(t^*+y)\geq D| S^*(t)\bigr) .
\]
The approximation in assumption A.2 is similar to the approximation
in the paper by \citet{LuMee93} which assumes that the probability
of a negative random slope in the linear model is negligible. One
particular case for the assumption A.2 to hold is that the signal is
monotone. However, monotonicity is not a necessary condition.
Assumption A.2 also holds for nonmonotone signals---an example of such
a signal is in Figure \ref{fig:exp:assumption:A2}.

Proposition \ref{prop2} below describes the updating procedure for RLD of a new
component given the posterior distribution of its
degradation process $S^*(\cdot)$ updated up to time $t^*$. The proof
follows directly as a consequence of Proposition \ref{prop1}.

\begin{prop}\label{prop2}
For a new partially degraded component
with its degradation process $S^*(\cdot)$ updated up to time $t^*$, the
residual life distribution is given as follows:
%
\begin{equation}\label{eq:distr}
P\bigl(T-t^*\leq y  | S^*(\cdot), T\geq t^*\bigr) =
\frac{\Phi_Z (g^*(y|t^*) ) -
\Phi_Z (g^*(0|t^*) )}{1-\Phi_Z (g^*(0|t^*) )},
\end{equation}
where $\Phi_Z$ represents the standard normal cumulative distribution
function and $g^*(y|t^*) =
\frac{\mu^*(t^*+y)-D}{\sqrt{V^*(t^*+y)}}$ with
\begin{eqnarray*}
\mu^*(t^*+y) &=& \mu(t^*+y) + (Cd)' p(t^*+y), \\
V^*(t^*+y) &=& \sum_{k_1=1}^K\sum_{k_2=1}^K
[C_{k1,k2}\phi_{k_1}(t^*+y)\phi_{k_2}(t^*+y) ].
\end{eqnarray*}
\end{prop}

In the above equations, $p(t^*+y)=(\phi_1(t^*+y),\ldots,\phi
_K(t^*+y))'$, and $C_{k1,k2}$ refers to the $(k_1,k_2)$ element of the
matrix $C$.

One advantage of obtaining the distribution rather than simply a
point estimate is that we can also derive a \textit{confidence
interval} for the remaining lifetime up to a degradation threshold
$D$. Following the derivation in Proposition \ref{prop2}, a $1-\alpha$
confidence interval for RLD is $[L,U]$ such that
\[
P\bigl(L\leq T-t^*\leq U| S^*(\cdot),  T\geq t^*\bigr) = 1-\alpha.
\]
Since we have one equation with two unknowns, the lower---$L$ and
the upper---$U$~tails are commonly equally weighted, and,
therefore,
\[
\frac{\Phi_Z (g^*(U|t^*) )-\Phi_Z (g^*(0|t^*)
)}{1-\Phi_Z (g^*(0|t^*) )}
= 1-\frac{\alpha}{2}
\]
and
\[
\frac{\Phi_Z (g^*(L|t^*) )-\Phi_Z (g^*(0|t^*)
)}{1-\Phi_Z (g^*(0|t^*) )}
= \frac{\alpha}{2}.
\]
However, we cannot obtain exact solutions for $L$ and $U$ because we
do not have a closed-form expression for the inverse of the
cumulative density function of $T-t^*$. For example, the first
relationship is equivalent to finding $U$ from $g^*(U|t^*) =
z_{\alpha_1}$ where $z_{\alpha_1}$ is the $1-\alpha_1$ quantile of
the normal distribution. Using this equation, we would like to
obtain $U$ such that
\[
\frac{\mu^*(t^*+U)-D}{\sqrt{V^*(t^*+U)}} = z_{\alpha_1},
\]
which is a nonlinear function of $U$ and its solution does not have
a close form expression. We therefore resort to \textit{parametric
bootstrap} [\citet{EfrTib93}; \citet{DavHin97}]
to sample from the distribution of $T-t^*$ which will give us a set
of realizations from this distribution---$T_1,T_2,\ldots,T_B$. Using
these realizations from the distribution of $T-t^*$, we estimate a
quantile bootstrap confidence interval.

The confidence interval estimation procedure is as follows. For
$b=1,\ldots,B$:
\begin{enumerate}
\item Sample $\xi^b = (\xi^b_1,\ldots,\xi^b_K)$ from the
multivariate normal distribution of the posterior scores provided
in Proposition \ref{prop1}.

\item Obtain a simulated signal
\[
S_b(t) = \mu(t)+\sum_{k=1}^{K} \xi_{k}^b \phi_k(t),
\]
where $\xi_{k}^b$, $ k=1,\ldots,K$, are the scores sampled at Step 1.

\item Take $T_b = \inf_{t\in[0,M]}\{S_b(t)\geq D\}$.
\end{enumerate}
Using the sampled values $T_1,T_2,\ldots,T_B$, we compute the
empirical $\alpha/2$ and $(1-\alpha/2)$ quantiles, $T_{\alpha/2}$
and $T_{1-\alpha/2}$, respectively. We estimate the upper and lower
bound of the confidence interval by $\hat L = T_{\alpha/2}$ and
$\hat U = T_{1-\alpha/2}$. It follows that $[\hat L,\hat U]$ is an
approximate $1-\alpha$ quantile bootstrap confidence interval for
the residual life time of the fielded component.

An additional approach to the (parametric) bootstrap method
described above is to (re)sample the signal data resulting in
multiple bootstrap samples. For each bootstrap sample, estimate the
residual lifetime using the approach discussed in this paper;
therefore, we obtain a set of realization from the distribution of
\mbox{$T-t^*$}. In contrast to the bootstrap method described above, this
alternative bootstrap approach requires estimating the FPCA model
for each bootstrap sample which is computationally expensive.

\section{Sampling scheme}\label{sec:expt:design}
The nonparametric degradation modeling framework
introduced in this paper applies to both complete as well as
incomplete degradation signals. For applications involving
incomplete degradation signals, it is important to develop a
sampling plan that ensures accurate estimation of the mean function
and the covariance surface. \citet{YaoMulWan05} provide theoretical
results on the estimation of the covariance surface using FPCA under
\textit{large $n$ but small $m_i$} for $i=1,\ldots,n$. In other words,
for these results to hold, the observation time points
$\{t_{ij}\}_{j=1,\ldots,m_i, i=1,\ldots,n}$ need to cover the time
domain, $[0,M]$, densely.

Using the traditional uniform sampling technique, the number of
observations per time interval decreases as more signals fail, leading
to an unbalanced number of observations per time interval---more
observations at the beginning of the observation time domain but fewer
observations at the end of the time domain. Further, this unbalanced
design will result in decreasing estimation accuracy (higher variances)
of the mean and covariance estimates at later time points. In order to
balance the number of observations per time interval throughout the
time domain $[0,M]$, we propose an experimental design using nonuniform
sampling. The proposed technique ensures relatively dense coverage of
the sampling time domain, $[0,M]$, where $M$ represents the last
observation time of the longest possible degradation signal for a given
application.

We note here that the sampling technique requires input of $M$ at the
beginning of the experiment although $M$ is unknown. It is often the
case that, in practice, an experimenter will set a timeline at the
beginning of the experiment which will specify a limit of how long the
experiment will be run (e.g., one year vs. one month). This upper limit
will specify $M$. Generally, starting with a lower initial value for
$M$ will allow the experimenter to sample densely enough while having
the option to update the sampling technique (update $M$) if not all
training signals have reached the failure threshold by the initial
value for $M$.

The following steps outline a sampling procedure for obtaining
sparsely observed and fragmented degradation signals:
\begin{longlist}[\textit{Step} 1:]
\item[\textit{Step} 1:] We begin by performing nonuniform sampling of the
time domain $[0,M]$, thus obtaining a sequence of time points,
$0=t_1<t_2<\cdots<t_{m-1}<t_m=M$, for large $m$. Since only a few
components will survive up to the maximum time point, $M$, we
increase the sampling frequency at later time points in order to
cover the sampling time domain at the extreme point, $M$.
Consequently, we sample exponentially, that is, the time interval
between two consecutive sampling time points decreases
exponentially over time (the decreasing rate is implicitly determined
by the value of $M$ and the number of sampling time points).

\item[\textit{Step} 2:] This step provides a potential sampling timetable
(or monitoring/\break observation schedule) for sparsely observed and
fragmented degradation signals. We begin by selecting $n$
components. For each component, we select its sampling time points
from the set $t_1,\ldots,t_m$ \textit{without any prior knowledge
about their degradation process and lifetime}. Next we define two
settings:
\begin{longlist}[Setting 1:]
\item[Setting 1:] This setting is used to obtain \textit{sparsely observed
degradation signals}. For component $i$, we randomly sample $m_i$
time points from the set of total time points
$\{t_1,\ldots,t_m\}$. This results in a set of sparse \textit{sampling
time points}, $\{t_{i1},\ldots,t_{im_i}\}$ for this
component.

\item[Setting 2:] This setting is used to obtain \textit{fragmented
degradation signals}. Recall that fragmented signals are obtained
by continuously monitoring a~component over a short time interval,
hence the term ``fragment.'' For component $i$, we select two or
more time points $B_1, B_2, \ldots$ from the set of total time
points $\{t_1,\ldots,t_m\}$. These points represent the beginning
times of the signal fragments or sampling intervals. The duration
of the sampling interval will depend on the type of application,
the availability of monitoring/testing equipment and the
associated costs/economics. Consequently, the end time points,
$E_1, E_2, \ldots,$ will vary from one experiment to another. In
other words, for component $i$, we may have two or more time
intervals: $[B_{i,1},E_{i,1}]$, $[B_{i,2},E_{i,2}]\ldots.$
\end{longlist}

\item[\textit{Step} 3:] Finally, we observe the degradation signal for the
selected components at the selected time points according to the
type of incomplete signals, sparse or fragmented, and obtain the set
of sampled signals $S_i(t_{ij})$ for $i=1,\ldots,n$ and
$j=1,\ldots,m_i$.
\end{longlist}

\begin{figure}

\includegraphics{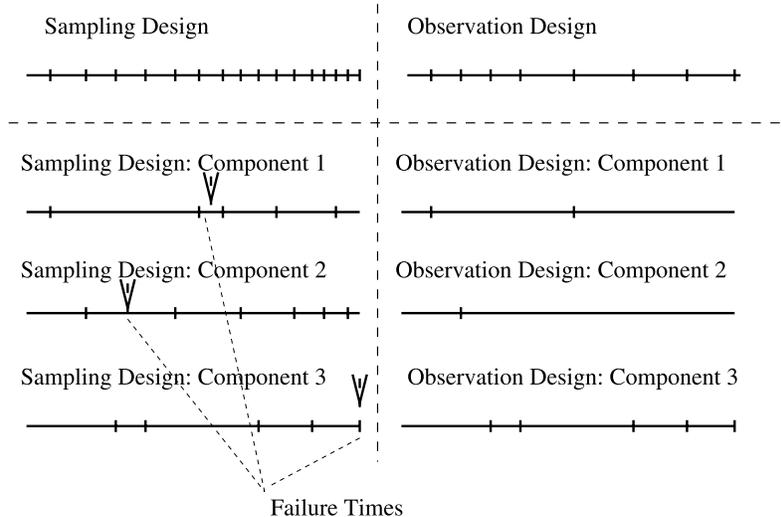}

\caption{Nonuniform sampling: Sampling time points vs.
observation time points.} \label{fig:exp:vs:obs}
\end{figure}

It is important to stress that we select the sampling time points
in Step~2 before observing the degradation signals. Since we do
not observe the failure time before selecting the time points, we
cannot ensure that the degradation signal will be observed for all
selected sampling time points. This is because some components may
fail before the latest selected time point. Consequently, the \textit
{observation time points} are a subset of the sampling time points
and will be less densely sampled close to $M$ since the missing
observations (the difference set between sampling and observation
time points) will increase in density closer to the upper bound
$M$. In Figure~\ref{fig:exp:vs:obs} we compare the sampling time
points selected at Step 2 to the observation time points for three
components. In this example the sampling time points are
nonuniformly selected, whereas the observation time points are
approximately uniform since for the first two components, we do
not observe at the latest times---only the third component fails
after its latest sampling time.

Two parameters that are used for tuning the sampling plan are as
follows: the
total number of sampling time points $m_{\mathrm{total}} = \sum_{i=1}^n
m_i$ and the total number of components $n$. The more sparsely the
signals are observed ($m_{\mathrm{total}}$ is small), the more signals we
need to observe ($n$ needs to be large). Selecting~$n$ and
$m_{\mathrm{total}}$ optimally is important to ensure accurate modeling of
the degradation process at a feasible cost. The larger the number
of components $n$ and/or the larger the number of time points
$m_{\mathrm{total}}$ are the higher the costs associated with monitoring
and testing. Note that selection of $n$ and $m_{\mathrm{total}}$ will vary
according to the type of application.

\section{Case study: Crack growth data}\label{sec:crack:study}

In this section we study crack growth data that can
be found in various domains of engineering applications, such as
infrastructure (bridges, steel structures), maritime (hulls of oil
tankers), aeronautical (aircraft fuselage), energy (vanes of gas
turbines), etc. We consider a situation in which crack growth data can
be observed from identical units (say, several ship hulls, or
turbines) up to a predetermined time period, denoted by $M$ in this
paper. A constant threshold, $D$, is a critical crack length
representing a soft failure when maintenance and repair should be
performed. Within this context, we assume that catastrophic failure,
that is, hard failure, may occur at a relatively larger crack length.

The data set used in our case study was first published in
\citet{VirHilGoe79}, and has been previously analyzed in other journal articles
[\citet{Kot98}; \citet{CroMakArm06} and the references therein].
The specimens in the test were 2.54-mm-thick and 152.4-mm-wide center
cracked sheets of 2024-T3 aluminum. The crack propagation signals of
these specimens were recorded under identical experimental conditions.
In this data set, the crack length was measured in millimeters and the
observation time was measured by the cumulative load cycles. More
details about this data set can be found in \citet{VirHilGoe79}. In
this study, we set the soft failure threshold to $D = 27$ mm. We
provide additional results for another soft threshold in the
supplemental material [\citet{ZhoSerGeb10}]. To be consistent with
the methodology in this study, the observations are censored at common
value $M=230\mbox{,}000$ cycles. A~representative example of sparsely sampled
degradation signals is in Figure \ref{fig::virkler}(a).

\subsection{Results and analysis}\label{sec:crack:study:sub2}

We report the prediction accuracy of the remaining life for varying
time points $t^*$ defining the latest observation time of a partially
degraded component. We consider the following degradation percentiles:
10\% (the signal has been observed up to time $t^*$, which equals to
10\% of the lifetime), 20\%, \ldots, 80\% and 90\%. For each crack, we
predict the updated residual lifetime at each of the nine percentiles
using the degradation signal observed up to that respective percentile.
The number of signals in this study is 59. We randomly select 50 of the
total signals as training signals for estimating the model components,
and the rest are validation signals for evaluating the performance of
our model in predicting residual life. For each validation signal, we
use the following error criteria to assess the prediction accuracy:
%
\begin{equation}\label{eq:error}
\mathrm{error}=\frac{|\mathrm{Estimated\ Life - Actual\
Life}|}{\mathrm{Actual\ Life}}.
\end{equation}
We replicate the above procedure for 100 times, and report the
distribution of the errors across the 100 simulations using a set of
boxplots, each boxplot corresponding to a degradation percentile for
the testing components and providing the absolute prediction errors for
that percentile.

\begin{figure}

\includegraphics{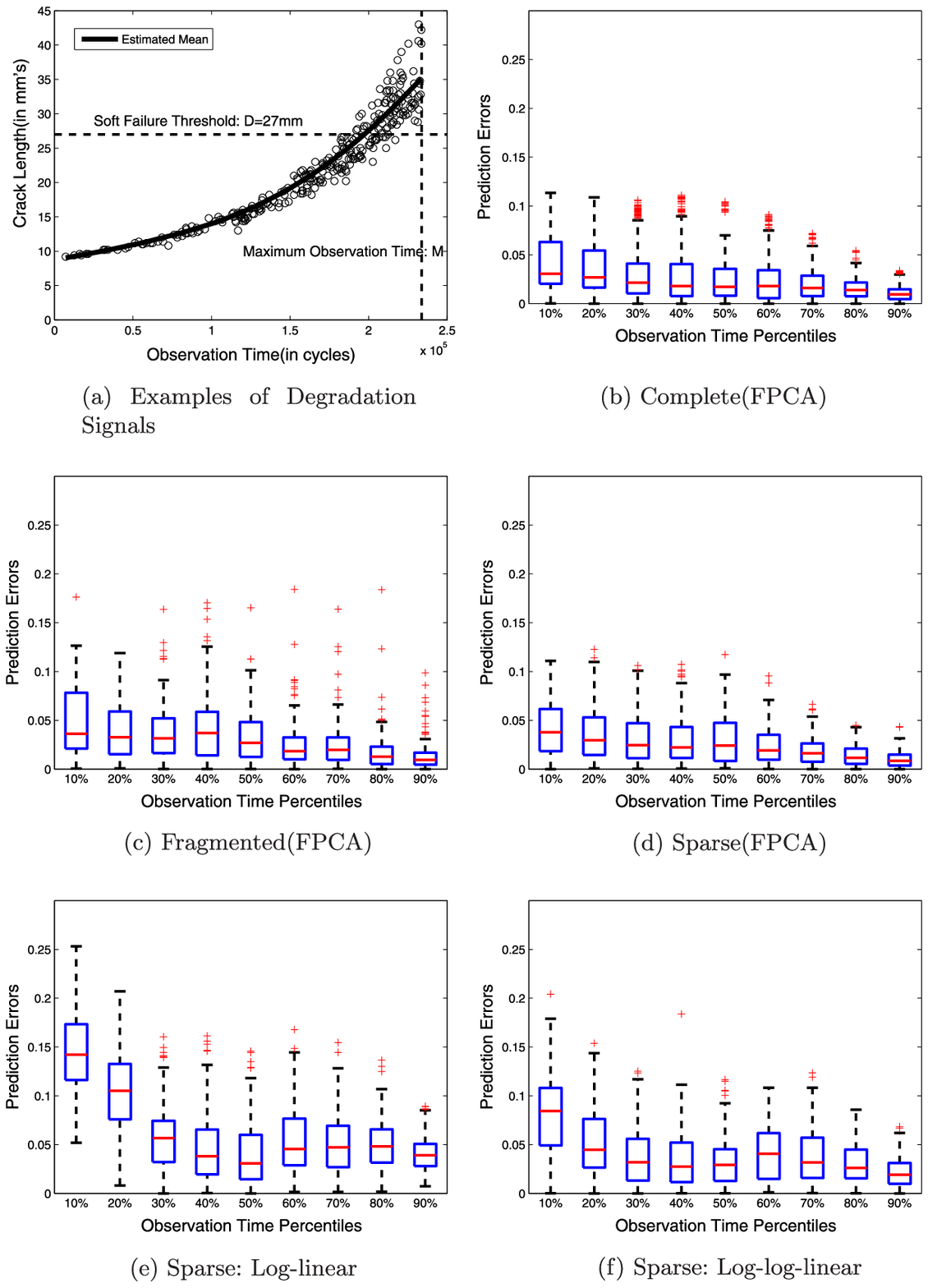}
\caption{The prediction error of residual life prediction for the
crack growth data set.}
\label{fig::virkler}
\end{figure}

We first discuss the performance of our nonparametric model
for complete, sparse and fragmented degradation signals. In
each complete degradation signal, we have about 50 observations per signal.
To obtain a sparsely observed degradation signal, we randomly
sample $m=6$ observations from each complete signal. We use two
intervals per signal to obtain fragmented degradation signals. The
results are illustrated in Figure {\ref{fig::virkler}}(b)--(d). The
results indicate that our nonparametric model performs well for
complete as well as incomplete degradation signals, and the
performance is better when the incomplete degradation signals are
sparse rather than fragmented. Although we have only approximately $10\%
$ observations of
complete degradation signals under the sparse sampling scenario, the
prediction errors do not increase
significantly. This observation is important in practice; under
budget limitations, one may resort to sparse or fragmented degradation
signals without significant loss of predictive capability.

We also demonstrate the benefits of our proposed nonparametric
degradation model by comparing it with parametric models as
benchmarks. Since the degradation signals have a nonlinear trend
with a curvature similar to the exponential function, we transform
the degradation signals using the natural logarithm in order to
linearize the trend and then apply a linear random effects
model (henceforth, denoted by ``log-linear''). Since under the
log-transform model, the residual life predictions are inaccurate
compared to the nonparametric approach, we consider a double
logarithm transformation of the degradation data (henceforth,
denoted by ``log--log-linear''). The results of the sparse scenario using
the parametric models ``log-linear'' and ``log--log-linear'' are
reported in Figure {\ref{fig::virkler}}(e)--(f), respectively. We find
that both
parametric models provide less accurate predictions of the
residual life than our nonparametric model. This is due to the
inaccuracy of
the parametric models in capturing the crack propagation trend.

\begin{figure}

\includegraphics{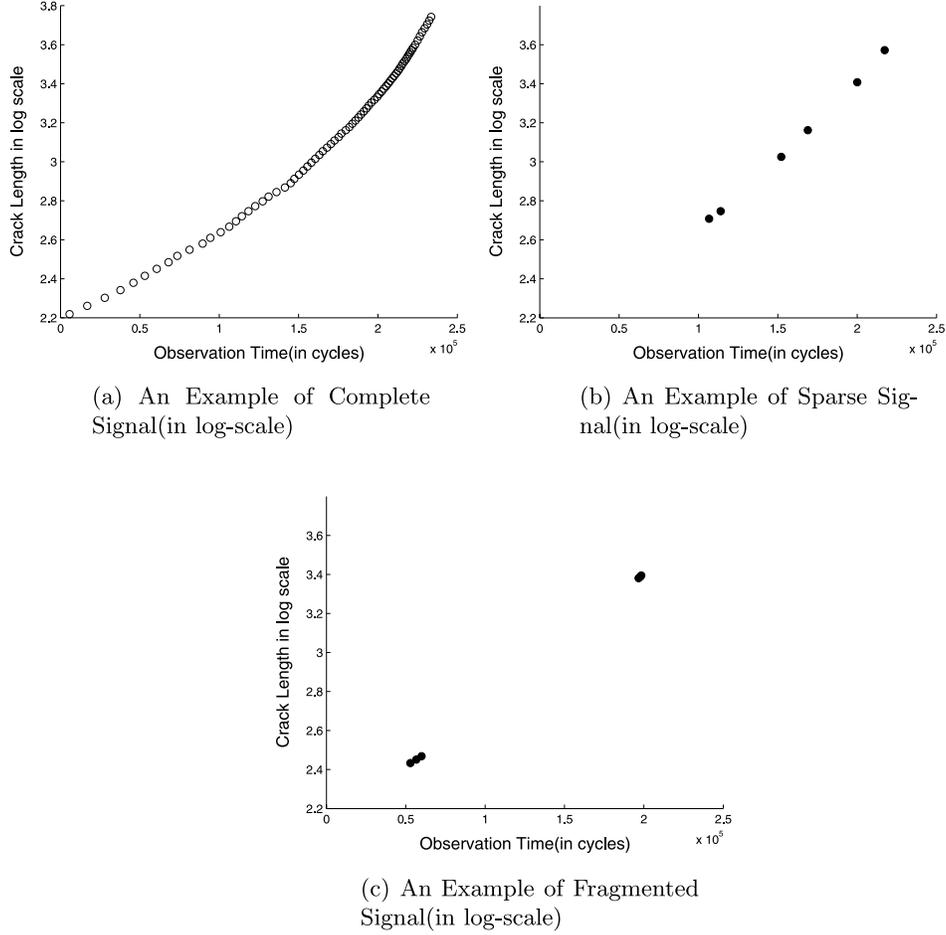}

\caption{Examples of the crack data under the log scale.}
\label{fig::virkler:log_bias}
\end{figure}

We provide one example in Figure \ref{fig::virkler:log_bias} to
illustrate the source of the bias of the ``log-linear'' model. In this
figure the $x$-axis represents the degradation time and the $y$-axis
represents the crack length, but in the log scale. We have one
complete, sparse and fragmented degradation signal in
Figu\-re~\mbox{\ref{fig::virkler:log_bias}(a)--(c)}, respectively. If the ``log-linear'' model
is the true underlying parametric model, we should see a linear trend
in all three plots. This seems to be true in the sparse or fragmented
cases [see Figure \ref{fig::virkler:log_bias}(b)--(c)]. However, for
Figure \ref{fig::virkler:log_bias}(a) showing a complete signal, we
note that the degradation trend is still nonlinear; the
log-transformation does not linearize the signal (the same applies for
the ``log--log'' transformation). Therefore, the ``log-linear'' model does
not accurately capture the crack propagation trend throughout the
unit's lifetime. This example shows the potential difficulty of
identifying a reasonable parametric model for sparse and fragmented
degradation signals and, in turn, demonstrates the robustness of our
proposed nonparametric model to model misspecification.

\section{Simulation study}\label{sec:simulation:study}
In this section we simulate nonlinear degradation
signals from three different models to demonstrate the benefits of
using our proposed nonparametric degradation modeling approach. We
evaluate our approach in terms of the prediction accuracy of
estimating the residual life for complete, sparse and fragmented
degradation signals, contrast uniform and nonuniform sampling
procedures for acquiring the ensembles of incomplete degradation
signals, and also investigate the robustness of our model to
violations of its model assumptions.










\subsection{Simulation models} The degradation
signals are simulated from three different models, and all of them
are special cases of the general model (\ref{eq:model}). More
specifically:

\begin{itemize}
\item In Model 1, we choose $\mu(t)=30 t^2$, $X_i(t)=\xi_{1} \phi
_1(t)$, where $\xi_1\sim N(0,\frac{45}{4})$,
$\phi_1(t)=\sqrt5t^2$, $0 \leq t \leq1$, and $\sigma=1$.

\item In Model 2, we choose $\mu(t)=30 t^2$, $X_i(t)=\xi_{1} \phi
_1(t)+\xi_{2} \phi_2(t)$, where $\xi_1\sim N(0,3^2)$, $\phi_1(t)=2t$, and
$\xi_2\sim N(0,(\frac{3}{2})^2)$,
$\phi_2(t)=\sqrt{80}t^2-\frac{3}{4}\sqrt{80}t$, \mbox{$0 \leq t \leq1$}.
(The coefficients of the eigenfunctions are chosen so that they form
an orthonormal functional basis for $0 \leq t \leq1$.)

\item In Model 3, we choose $\mu(t)=30 t^2-2\sin(4 \pi t)$, $X_i(t)=\xi
_{1} \phi_1(t)+\xi_{2} \phi_2(t)$, where $\xi_1\sim N(0,3^2)$, $\phi
_1(t)=2t$, and
$\xi_2\sim N(0,(\frac{3}{2})^2)$,
$\phi_2(t)=\sqrt{80}t^2-\frac{3}{4}\sqrt{80}t$, $0 \leq t \leq1$.
\end{itemize}

We simulate from Model 1 because its residual life distribution
can be easily derived from training signals and updated using
validation signals using the procedure in
\citet{Gebetal05}. The derived residual life distribution can then be utilized as
a benchmark to assess the performance of our nonparametric
approach.

Across all the models, the failure threshold is set to $D=10$. We
generate $n=100$ ``training'' signals and $n=100$
``validation'' signals from each model. For a complete signal, we have 51
observations made at an equally spaced grid ${c_0, \ldots, c_{50}}$ on
$[0,1]$ with $c_0=0, c_{50}=1$. A sparse or fragmented signal is
then sampled from a complete signal such that we observe
about 6 observations per signal. The stopping time for each training
signal (the last point at which a signal is observed) is generated
from Uniform distribution [$\operatorname{Uniform}(0.7,1)$]---our simulation
results are insensitive to the selection of the stopping time
distribution.

We run simulations for $100$ times. For each simulation, we compute
the prediction errors at the following degradation percentiles: 10\%,
20\%, \ldots, 70\%, 80\% and 90\% of the simulated degradation signals.


\begin{figure}
\begin{tabular}{@{}cc@{}}

\includegraphics{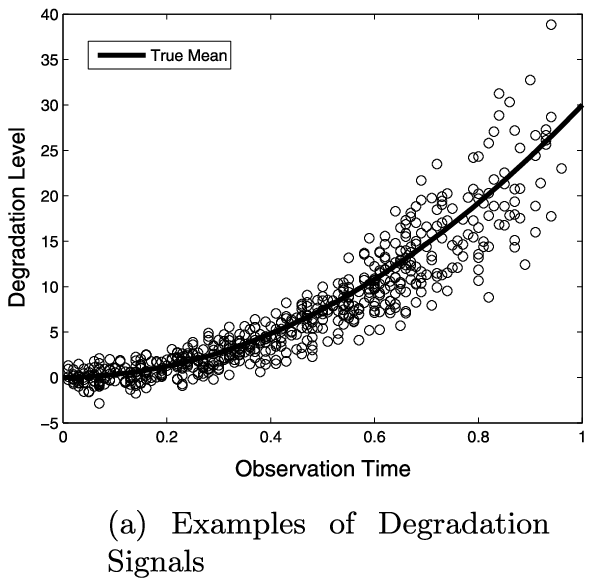}
&\includegraphics{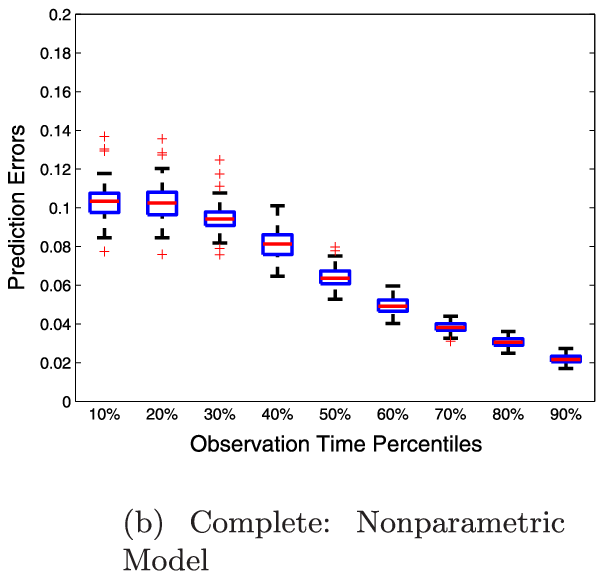}\\

\includegraphics{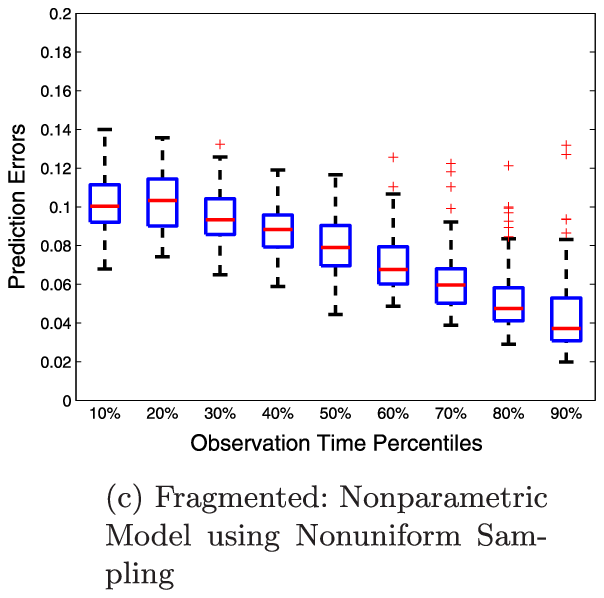}
&\includegraphics{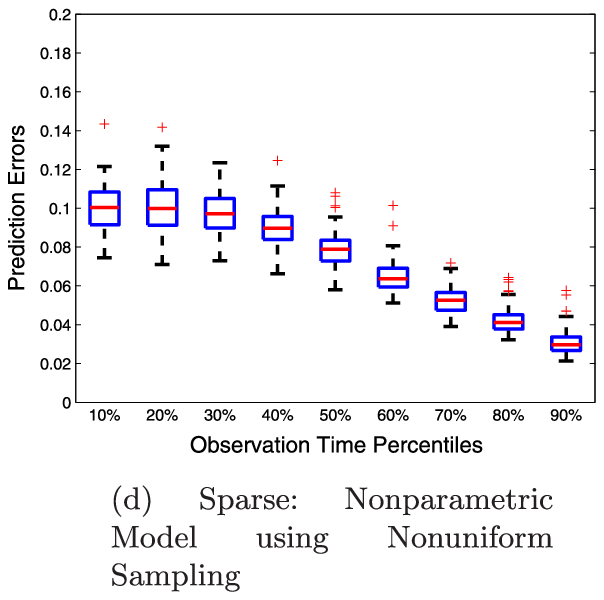}\\

\includegraphics{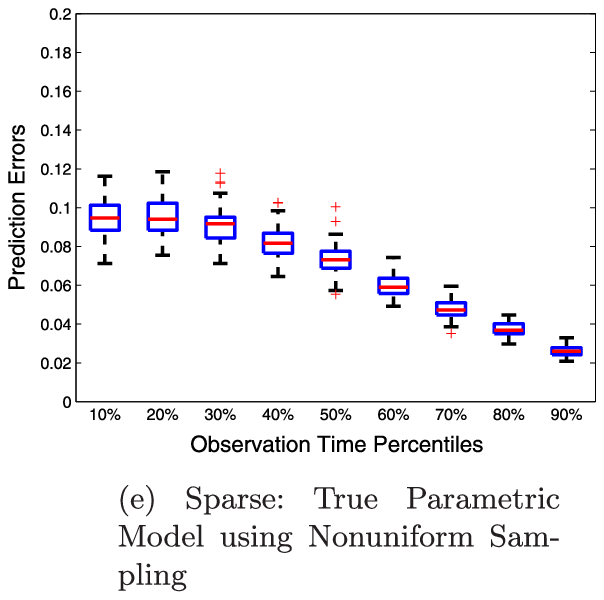}
&\includegraphics{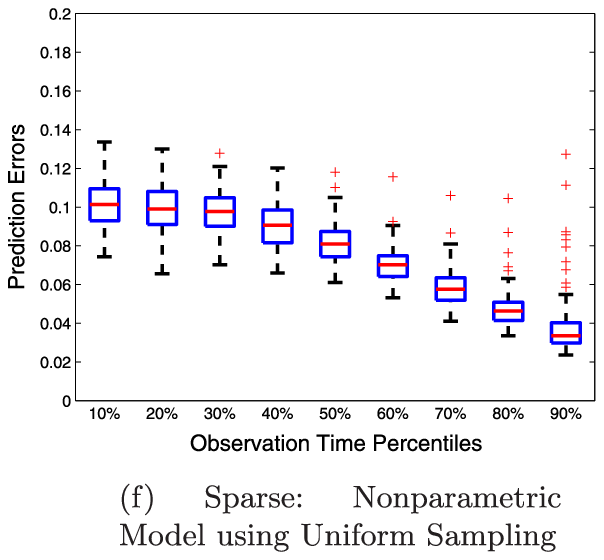}
\end{tabular}
\caption{The prediction error of the residual life estimate for
Model 1.}
\label{fig::sim:results1}
\end{figure}
\setcounter{figure}{5}
\begin{figure}
\begin{tabular}{@{}cc@{}}

\includegraphics{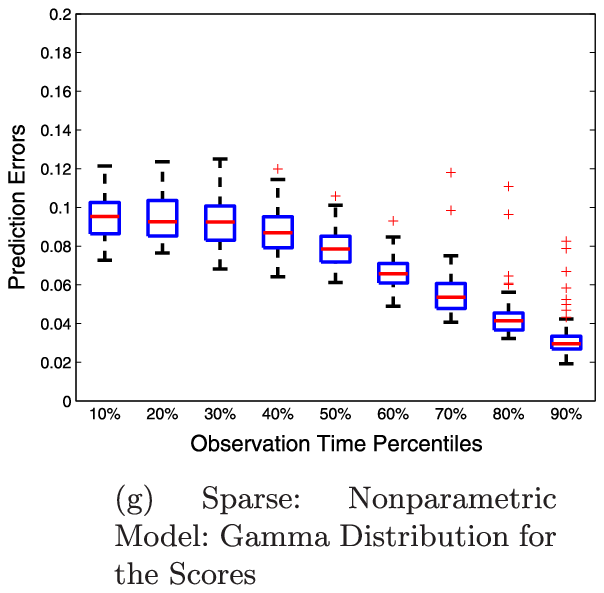}
&\includegraphics{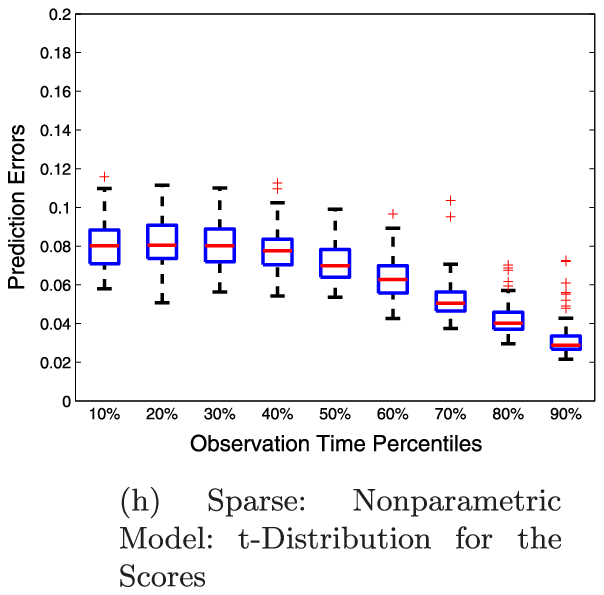}\\
\multicolumn{2}{@{}c@{}}{
\includegraphics{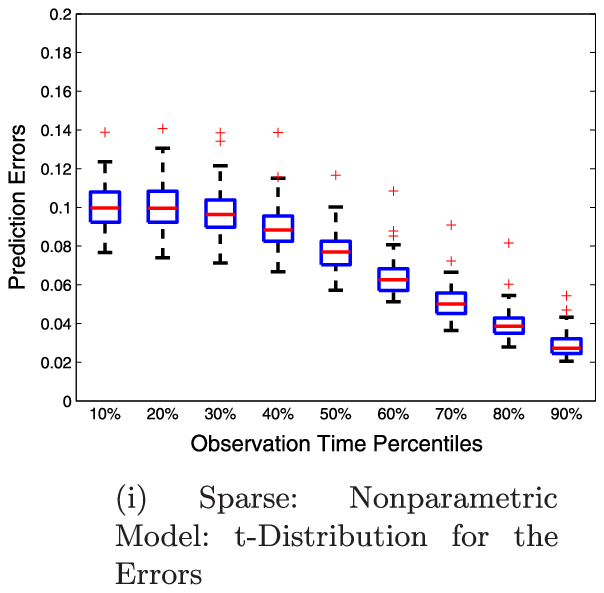}
}
\end{tabular}
\caption{(Continued).}
\end{figure}

\subsubsection{Results and analysis of Model 1}

In Figure \ref{fig::sim:results1}(b)--(d), we present the boxplots of the
prediction
errors when using the nonparametric degradation model in this paper for
complete, fragmented and sparse degradation signals. For the sparse
scenario, we compare the prediction accuracy of using the true
parametric model [see Figure \ref{fig::sim:results1}(e)] and our
nonparametric model when signals are
uniformly sampled [see Figure \ref{fig::sim:results1}(f)] or
nonuniformly sampled [see Figure \ref{fig::sim:results1}(d)]. We assess
the robustness to model assumptions by simulating signals from the
model with $\xi_{1}$ following a Gamma or Student~$t$ distribution [see
Figure \ref{fig::sim:results1}(g)--(h)]. We also compute the prediction
errors under different error distributions [see Figure \ref
{fig::sim:results1}(i)].

The first observation is that there is insignificant difference in the
prediction errors between the true parametric model and the
nonparametric degradation model. The differences are larger for
high degradation percentiles. Since the difference in the prediction
errors increases with additional data, we observe for a new
component, we infer that this small inefficiency arises due to a
decreased accuracy in the estimation of the empirical prior
distribution at the later time points.

The second important observation is that the nonuniform sampling
technique proposed in Section \ref{sec:expt:design} enhances the
prediction accuracy of the residual life. In Table \ref
{table:nonuniform:decrease} we list the median prediction errors based
on nonuniform sampling and uniform sampling techniques. The first row
of this table represents the time percentile of the degradation signals
used for predicting the residual life. It is apparent that the
nonuniform sampling technique provides smaller prediction errors,
especially at high time percentiles. This is because nonuniform
sampling ensures dense coverage of observations over the whole time
domain, including the region near maximum observation time ($M$), and
hence provides more accurate estimate of the mean and covariance
functions of the model, especially at higher time percentiles.

\begin{table}[b]
\caption{Prediction errors based on sparse degradation signals that are
uniformly or nonuniformly sampled}\label{table:nonuniform:decrease}
\begin{tabular*}{\tablewidth}{@{\extracolsep{\fill}}lcccccccc@{}}
\hline
\textbf{Time percentiles} & \textbf{20\%} & \textbf{30\%} & \textbf{40\%} & \textbf{50\%} & \textbf{60\%} & \textbf{70\%} & \textbf{80\%} &
\textbf{90\%} \\
\hline
Uniform &10.08 & 9.75 &9.01& 8.17 &6.91& 5.77& 4.79& 3.95 \\
Nonuniform &10.08 & 9.75 &8.97& 7.89 &6.50& 5.28& 4.23& 3.11\\
\hline
\end{tabular*}
\end{table}

Last, we assess the robustness to departures from our model
assumptions: normality of the scores and normality of the errors. In
Figure \ref{fig::sim:results1}(g)--(h), we compare the
prediction errors when the scores follow Gamma and Student~$t$
distribution. We also present the results when the errors follow
Student~$t$ distribution in Figure~\ref{fig::sim:results1}(i). The
prediction errors for all
these different settings are similar. This robustness property of
our degradation modeling is inherited from the robustness of the
FPCA method [\citet{YaoMulWan05}].

\begin{figure}

\includegraphics{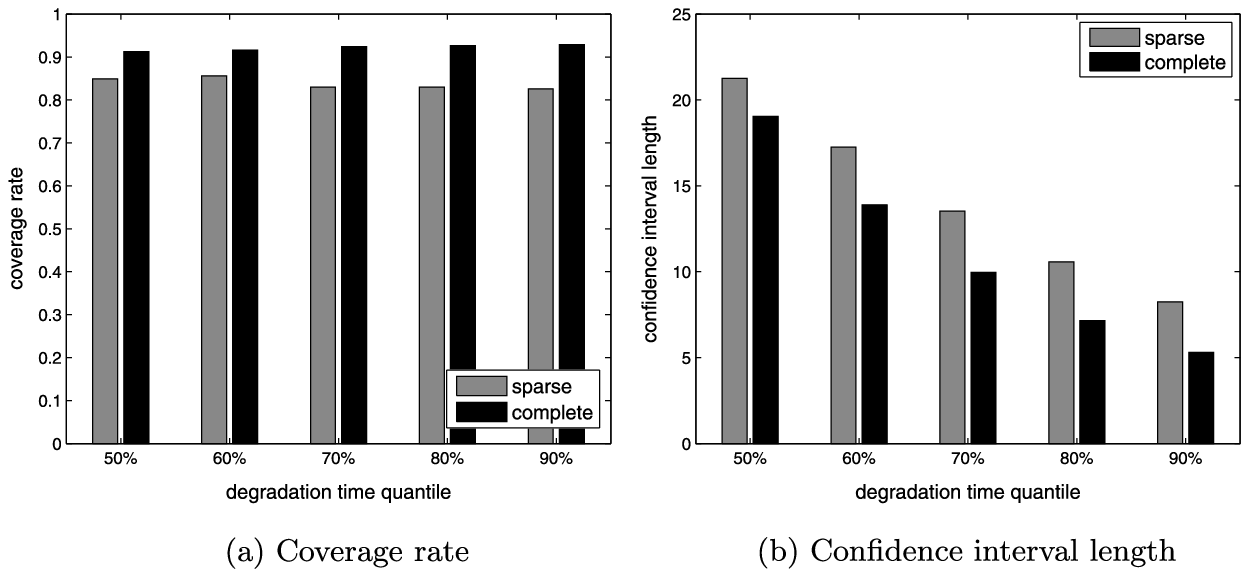}

\caption{Confidence interval estimation: the coverage rate \textup{(a)} and mean
length \textup{(b)}. In each plot the left and the right bars correspond to the
sparse and complete scenarios, respectively.} \label{fig::sim:ci}
\end{figure}

We also evaluate the accuracy of the confidence interval estimates
introduced in Section \ref{sec:lifetime:distr}. In Figure
\ref{fig::sim:ci} we present the coverage rate level and the mean
of the confidence interval length at the degradation lifetime
percentiles 50\%, 60\%, 70\%, 80\% and 90\%. The confidence interval
level is $1-\alpha= 0.9$. The coverage rate is higher for complete
signals than for sparse signals throughout all percentiles, but the
difference is insignificant. The coverage rate for both complete and
sparse signals is approximately equal to the confidence level
$1-\alpha= 0.9$. Moreover, the mean length decreases for higher
percentiles, implying that the accuracy of the residual life estimate
increases as the latest observation time point $t^*$ is closer to
the failure time.

\subsubsection{Results and analysis of Model 2}

In the following analysis we still use Model 1 as the assumed
parametric model and its derived residual life distribution as the
benchmark. This assumed parametric model correctly captures the mean
degradation trend of Model 2 but not the underlying covariance
structure of the degradation process. It is worth mentioning that
most existing parametric approaches focus on identification of the
functional form for the underlying degradation trend, ignoring the
underlying covariance structure.

\begin{figure}

\includegraphics{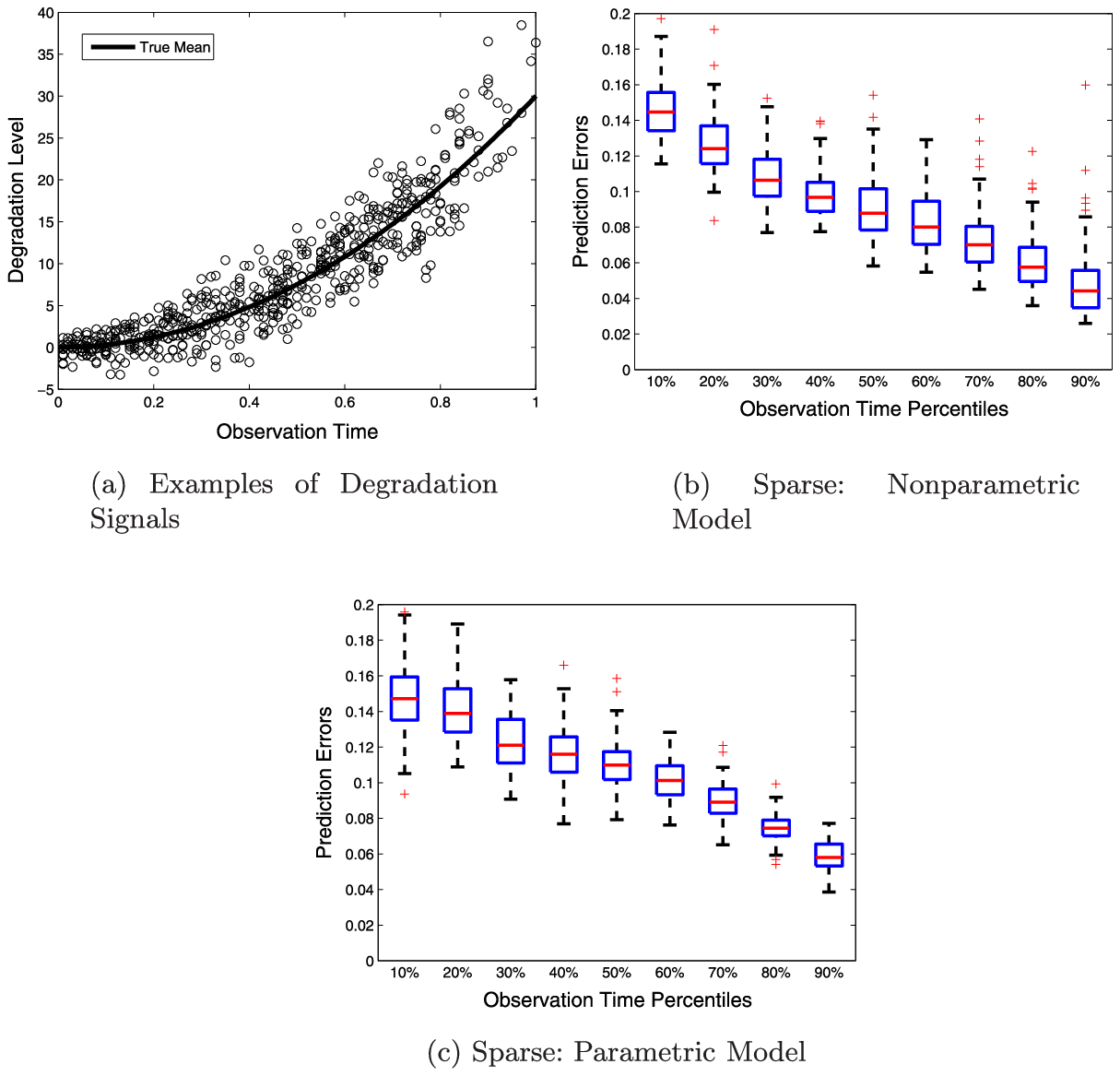}

\caption{The prediction error of the residual life estimate for
Model 2.}
\label{fig::sim:results2}
\end{figure}

The results in Figure \ref{fig::sim:results2} indicate that our
nonparametric model is more accurate than the assumed parametric
model in predicting the residual life. This is because our proposed
nonparametric approach, which is FPCA-based, cannot only estimate
the mean trend accurately but also capture the dominant modes of the
covariance structure correctly. In contrast, parametric models are
not flexible enough to accurately capture the underlying covariance
structure.

We also compute the prediction error results for cases when the
observed degradation signals are complete, fragmented or sparse, and
also when the scores and errors follow different distributions.
Detailed results can be found in the
supplementary materials [\citet{ZhoSerGeb10}].

\subsubsection{Results and analysis of Model 3}

We discuss this in the supplementary materials [\citet{ZhoSerGeb10}].

\section{Discussions}

In this paper we propose an Empirical Bayesian method for
predicting the degradation of a partially degraded component or
system. Specifically, we assume that the degradation process has
unknown mean and covariance, which can be estimated through a
nonparametric approach using a historical database of degradation
signals used to estimate the prior distribution of the degradation
process. These training degradation signals may be completely or
incompletely observed, that is, may be in the form of sparsely observed
signals or fragmented signals.

Our degradation modeling and monitoring approach relies on a series of
assumptions:

\begin{itemize}
\item The degradation signals follow a Gaussian process.

\item The time points at which the training signals have been observed
cover the time domain $[0,M]$ cumulatively.

\item The degradation signal of the new component does not cross back
the threshold~$D$.
\end{itemize}

From our simulation results, departures from the Gaussian assumption
will insignificantly alter the residual life estimates when a large
number of training signals are observed, as discussed in Section
\ref{sec:simulation:study}. This property is inherited from the
robustness of the FPCA approach used in estimating the empirical
prior distribution.

Under sparse sampling, the selection of the observation times of the
training degradation signals impacts the accuracy of the degradation
prior modeling. For example, if the degradation signals are
uniformly but sparsely sampled, the degradation process will not be
adequately observed at the later extreme time point $M$, since few
components will survive up to this time point. Consequently, uniform
sampling compromises the accuracy of the mean and covariance
estimates of the prior degradation process, which, in turn,
compromises the accuracy of the residual life estimate. In the
simulation study we show that the accuracy of the residual life
estimates is low for the traditional uniform sampling as compared to
the accuracy of the estimates under nonuniform sampling. Thus, the
second assumption is ensured under nonuniform sampling but not
uniform sparse sampling (see Section~\ref{sec:expt:design}).

The third assumption in our modeling approach relies on that the
experimenter will shut off or replace the component shortly after it
degraded beyond the failure threshold $D$.

In this paper we have applied the nonparametric approach to crack
growth data with a wide applicability, for example, in infrastructure
(bridges, steel structures), maritime (hulls of oil tankers),
aeronautical (aircraft fuselage), energy (vanes of gas turbines) and
others. This case study demonstrates the accuracy of the
nonparametric approach introduced in this paper as compared to
random effects parametric models which impose constrains on the
shape of the trend $\mu(t)$ and the covariance $C(t,t')$. Other
potential applications are relevant to LED data that could be found
in \citet{YuTse98}, \citet{LiaTse06} and \citet{TsePen07}.

\section*{Acknowledgments}
We would like to thank the Editor, anonymous
Associate Editor and anonymous reviewers for their constructive and thoughtful
comments on this manuscript.


\begin{supplement}
\stitle{Additional results}
\slink[doi]{10.1214/10-AOAS448SUPP} 
\slink[url]{http://lib.stat.cmu.edu/aoas/448/supplement.pdf}
\sdatatype{.pdf}
\sdescription{In this supplemental file we provide some additional
results of the crack growth data study and the simulation study.}
\end{supplement}

%

\printaddresses

\end{document}